\documentclass[letterpaper,twocolumn,10pt]{article}
\usepackage{usenix-2020-09}
\usepackage[utf8]{inputenc}
\usepackage{array, graphics, graphicx, color, xurl, xspace, multirow, rotating, epsfig, amsmath, amsfonts, booktabs, enumitem,sidecap,authblk}

\makeatletter
\renewcommand\AB@affilsepx{\hfill \protect\Affilfont}
\makeatother

\title{Building Flexible, Low-Cost Wireless Access Networks With Magma}
\author[1]{Shaddi Hasan}
\author[2]{Amar Padmanabhan}
\author[3]{Bruce Davie}
\author[4]{Jennifer Rexford}
\author[5]{Ulas Kozat}
\author[5]{Hunter Gatewood}
\author[5]{Shruti Sanadhya}
\author[5]{Nick Yurchenko}
\author[5]{Tariq Al-Khasib}
\author[5]{Oriol Batalla}
\author[5]{Marie Bremner}
\author[5]{Andrei Lee}
\author[5]{Evgeniy Makeev}
\author[5]{Scott Moeller}
\author[5]{Alex Rodriguez}
\author[5]{Pravin Shelar}
\author[5]{Karthik Subraveti}
\author[5]{Sudarshan Kandi}
\author[5]{Alejandro Xoconostle}
\author[5]{Praveen Kumar Ramakrishnan}
\author[6]{Xiaochen Tian}
\author[5]{Anoop Tomar}

\affil[1]{Virginia Tech}
\affil[2]{Databricks}
\affil[3]{Systems Approach}
\affil[4]{Princeton}
\affil[5]{Meta}
\affil[6]{Independent}
\date{} %

\newenvironment{CompactItemize}
  {\begin{itemize}[noitemsep,topsep=0pt]}
  {\end{itemize}}

\newcommand{\ra}[1]{\renewcommand{\arraystretch}{#1}}

\newcommand{\orctr}{orchestrator\xspace}
\newcommand{\name}{Magma\xspace}
\newcommand{\ffi}{FreedomFi\xspace}
\newcommand{\aparks}{AccessParks\xspace}

\newcommand{\drop}[1]{}
\newcommand{\allnotes}[1]{}

\begin{document}
\thispagestyle{empty}
 \maketitle

\begin{abstract}
\em
Billions of people remain without Internet access due to availability or affordability of service.
In this paper, we present \name, an open and flexible system for building low-cost wireless access networks.
\name aims to connect users where operator economics are difficult due to issues such as low population density or income levels, while preserving features expected in cellular networks such as authentication and billing policies.
To achieve this, and in contrast to traditional cellular networks, \name adopts an approach that extensively leverages Internet design patterns, terminating access network-specific protocols at the edge and abstracting the access network from the core architecture.
This decision allows Magma to refactor the wireless core using SDN (software-defined networking) principles and leverage other techniques from modern distributed systems.
In doing so, \name lowers cost and operational complexity for network operators while achieving resilience, scalability, and rich policy support.

\end{abstract}

\graphicspath{ {./images/} }
\section{Introduction}
\label{sec:intro}
Good Internet connectivity has become a basic necessity for people and enterprises all over the world.  Yet, more than one-third of the global population does not have access to the Internet~\cite{Internetstats}, and many other users do not have the high-speed connectivity needed for many important applications.  
The problem is primarily a matter of economics: commercial network operators claim that today's Internet has reached the user footprint that seems commercially viable to serve~\cite{gsma-unlocking}. 
To reach the next billion users, we must reduce the \emph{cost} of providing Internet access or enable actors beyond traditional, large-scale commercial operators to build sustainable, scalable network infrastructure.  
We need effective ways to reduce both capital and operational costs, through less expensive equipment and software, less reliance on highly skilled network administrators, and increased utilization of existing local capabilities.
At the same time, providers need ways to manage their limited network resources effectively 
to enable sustainable network operation.
Cellular networks typically achieve these goals with per-user policies, which may include per-user data caps, rate limits, or usage-based charging.

Unfortunately, conventional wireless solutions are not well suited to many scenarios affecting 
under-served users. WiFi access points operating on unlicensed spectrum cannot generally provide efficient coverage to large geographic regions (e.g., sparsely populated rural areas) due to the propagation characteristics of the radios. Plus, WiFi networks typically do not offer fine-grained policies to manage resources. In contrast, cellular base stations offer wider coverage, support more users, and connect to core networks that support more flexible policies. 
However, today's cellular access networks rely on expensive equipment, complex protocols, and a highly skilled workforce, limiting their ability to cost-effectively connect the next billion. While cellular networks scale \emph{up} to large user populations, they do not scale \emph{down} well. That is, a small cellular deployment is typically quite expensive. Magma aims to bridge the gap between these two classes of solution: cellular networks with rich policies, large user
populations, and long distances, 
and the simpler but less scalable WiFi networks.

More fundamentally, we observe that choosing to use a cellular radio access network (RAN) today \emph{forces} a network operator to make a series of decisions that deeply impact their network operations that are not 
inherently related to their choice of access network technology.
This choice binds a network operator to:
(i) a specific network architecture---namely the 3GPP-defined arrangement of interfaces for network management and on-path devices for policy enforcement, 
(ii) an ecosystem of vendors that has largely evolved to meet the needs of massive-scale telecom operators, and
(iii) a particular set of radio frequencies and associated regulatory requirements.
\noindent
The Magma project aims to change all this, by creating an open-source, carrier-grade wireless networking platform that supports a wide range of deployment scenarios.  
Magma deployments can leverage whatever radio access technology is readily available and most appropriate for their density of subscribers or deployment scenario. 
Magma achieves this goal through \emph{access gateways} that terminate the radio-specific protocols as close to the radios as possible. As a result, Magma allows carriers to augment an existing cellular deployment with WiFi hotspots in popular locations (e.g., athletic venues), or use LTE base stations to serve homes in rural areas, using a single core network and management platform.  

Ideally, new deployments could start small and grow over time.  
Magma achieves a ``scale as you go'' design through horizontal scaling of software components that run on commodity hardware, as is common in cloud-computing environments.  Magma also leverages  open-source software components
(e.g., Open vSwitch, gRPC, Kubernetes, Prometheus) 
commonly used in cloud settings.
Magma simplifies network management by adopting software-defined networking concepts, so that a central point of control can be used to set network policies, manage subscribers, etc. Magma adopts a hierarchical control plane
to improve scalability. 
\name supports only the essential features for efficient Internet access (e.g., authentication, accounting, and per-user policies), and forgoes some complex features.  For example, while \name supports both mobility (within the area served by an access gateway) and roaming, it does not yet support seamless user mobility {\em between} access gateways; this is because mobility has not been a requirement for the use cases 
that commercial 
deployments of \name support (e.g., home broadband or backhaul to WiFi hotspots). 
As other work has observed~\cite{cellbricks}, modern end-host protocols and applications can perform well without in-network mobility support.

In this experiences paper, we present the lessons learned in designing and deploying Magma.  We discuss how the goals of supporting heterogeneous 
radio and backhaul 
technologies and flexible policies, all at low cost, lead to a novel software architecture. Magma is used in real-world  deployments that vary significantly in geographic scope, number of users, technology choices, and the business models
that make them financially sustainable.
In Section~\ref{sec:backg} we motivate Magma's central tenet that the radio access technology should not dictate the network architecture.  Then, Section~\ref{sec:arch} discusses how the design of the access gateways enables Magma to support diverse technologies, a scalable control plane, fault tolerance, and more.  
Next, Section~\ref{sec:eval} presents an experimental evaluation that demonstrates that Magma 
design and implementation 
achieves good performance and scalability along with a discussion of two production access networks. 
We have seen cost savings in one deployment of 43\% compared to traditional approaches due to lower operational, hardware, and software costs. Our deployment experience also illustrates how \name scales both up and down, with one deployment supporting more than 800 eNodeBs (base stations) in 45 US states at the time of writing.
Section~\ref{sec:related} presents related work.
The paper concludes in Section~\ref{sec:concl} with a discussion of ongoing work on Magma and future challenges.

\textbf{Ethics.} This paper raises no ethical concerns. For the deployments discussed in Section~\ref{sec:eval}, we only consider operational data and did not have access to any user data or traffic.

\section{The Radio Access Technology Should Not Drive the Network Architecture}
\label{sec:backg}
Traditionally, the choice of radio access technology dictates a raft of other decisions about the network architecture. 
In contrast,  Magma  starts with the premise that each radio access technology has a role to play in reaching diverse user communities and that network operators should be able to use the radio and backhaul technologies most suited for a deployment scenario.
In short, wireless network architectures should, like the Internet itself, abstract away the link layer.

\subsection{WiFi vs. Cellular Access Networks}
The two main classes of radio access technologies emerged as extensions to existing wireline networks with different design philosophies.
WiFi extended IP networks, whereas modern cellular data networks began as extensions to voice telephony networks. Many of the differences between these two classes of access networks follow directly from this early distinction.

\textbf{WiFi:} WiFi allows inexperienced users to run simple low-cost local-area networks on their own. 
These networks use unlicensed radio spectrum (typically at 2.4 GHz and 5 GHz) that do not require WiFi network operators to get advance regulatory approval. 
At the same time, anyone can access the same spectrum, subject to limits on transmission power.
As a result, WiFi networks share their bands with devices including baby monitors, cordless phones, and smart power meters, so the WiFi MAC layer must assume that a WiFi access point (AP) operates in the presence of physical-layer interference. 
Combined with power restrictions that limit transmit distance, WiFi is most suitable for dense coverage in small areas.
WiFi service is best-effort, consistent with the Internet design philosophy---and realistic given the likelihood of interference. 
Enterprise WiFi deployments, such as those on college campuses and in corporate office buildings, perform more centralized management of interference across multiple overlapping access points.  Still, the risk of interference means that the service remains best-effort. 

\textbf{Cellular:} Cellular access networks allow telecommunication providers to offer wireless service to their subscribers, typically using licensed spectrum that is owned or leased by the carrier for long periods of time at high cost.
Since the radio has exclusive access to spectrum over a geographic region,
cellular waveforms are designed for wide-area coverage and high spectral efficiency, with deployments by well-resourced actors that can acquire land, build and connect towers, and hire skilled staff.

Regardless of access technology, any network of significant scale requires substantial investment in equipment, staffing, and, in the case of cellular networks, regulatory licenses. 
Thus, beyond very small networks, operators implement policies to manage limited spectrum, ranging from access control; charging for service based on time, usage, or more sophisticated techniques that incorporate community values~\cite{johnson2021commgestion}; usage caps; and throttling. The policy specification for LTE, for example, runs to almost 300 pages~\cite{3GPP-policy}. A simple example policy would be: ``rate limit customer C to X Mbps until they have sent Y GB in interval $t_{1}$, then limit to Z Mbps for interval $t_{2}$.'' 
Supporting flexible policies can help carriers reach under-served users in a financially sustainable manner: even networks operating for %
social reasons still incur costs 
and must efficiently manage limited resources.

These capabilities are implemented by a sophisticated packet core network that connects multiple base stations to the Internet.  
In contrast to how the Internet architecture changes incrementally, each generation of cellular network has been an opportunity to rethink everything from authentication to the modularity of the control and data planes.  

\begin{figure}
\centering
\includegraphics[width=0.99\columnwidth]{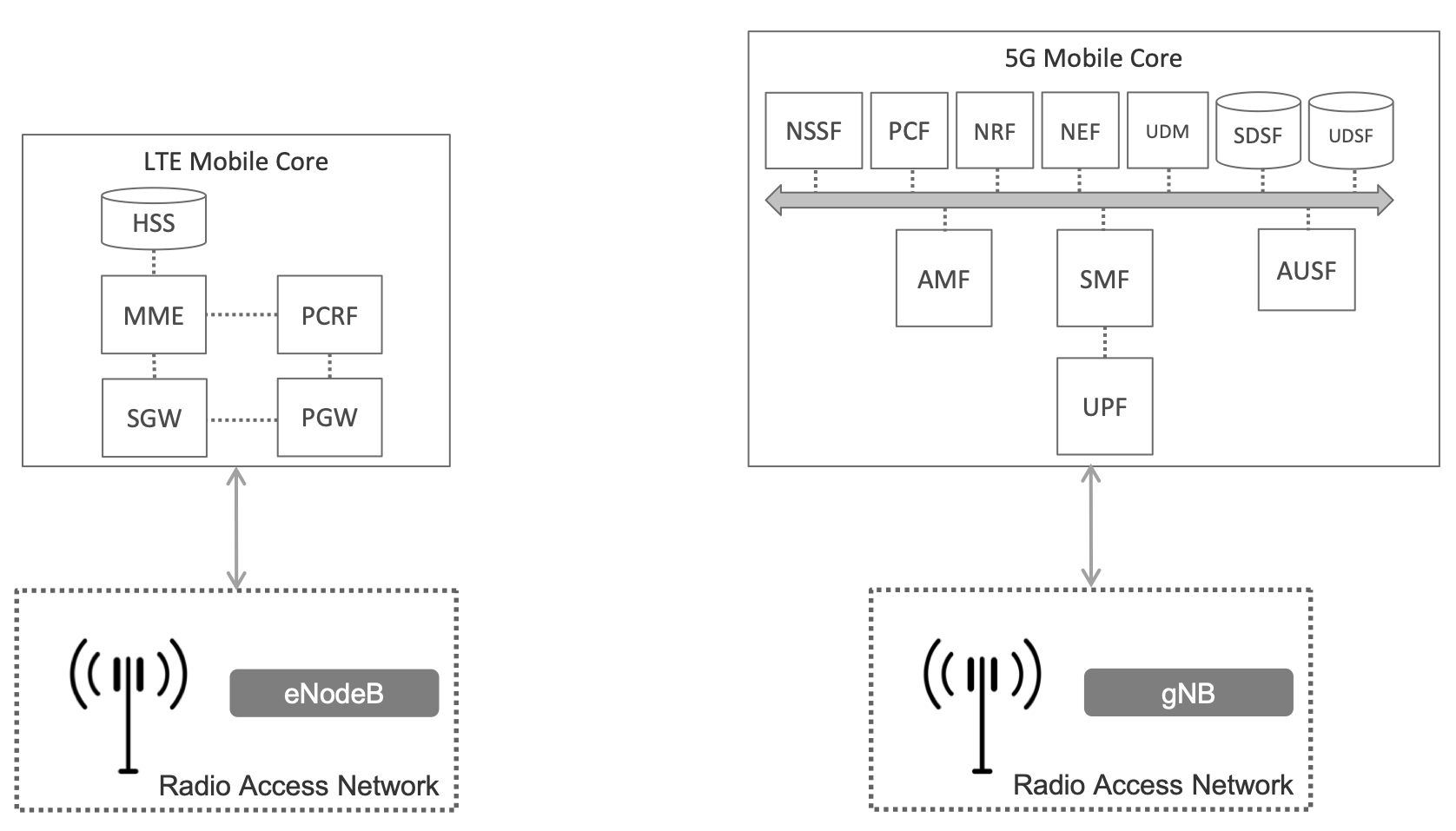}
\caption{Differences between the LTE and 5G architectures. See Appendix for explanation of acronyms.}\label{fig:4g-5g}
\vspace{-0.1in}
\end{figure}

As such, different generations of the 3GPP standards~\cite{www-3gpp} have different packet core architectures. UTMS (``3G'') differs from LTE (``4G''), which differs from 5G, and all of the generations differ from enterprise WiFi. 
The differences between LTE and 5G are illustrated in Figure~\ref{fig:4g-5g}, adapted from \cite{5G:SA}. The different radio technologies require differences in the base stations (eNodeB versus gNB) but note also the change in modularity of the mobile core. 
WiFi would be different again, and less standardized, with functions such as Authorization, Authentication, and Accounting (AAA) corresponding roughly to Mobility Management Entity (MME) and Home Subscriber Server (HSS) components in LTE.

Today, the boundaries between cellular and WiFi are increasingly blurry, with operators deploying each technology in scenarios more classically served by the other.
In recent years, large WiFi deployments have adopted more sophisticated methods for user authentication, power control, seamless mobility, and more~\cite{wifi-qos,wifi-handover, wifi-security}, with efforts like Eduroam~\cite{eduroam} and OpenRoaming~\cite{openroaming} bringing cellular-like 
wide area 
roaming to users of WiFi access networks.
Similarly, some cellular access networks now use ``lightly licensed'' spectrum, such as Citizen's Band Radio Service (CBRS)~\cite{cbrs} that supports dynamic allocation of radio spectrum to give radios exclusive access to some portion of the spectrum (on the timescales of tens of minutes).
Enterprises are deploying \emph{private} cellular access networks for a range of use cases---such as industrial automation, medical applications, and Internet access at hotels and sporting events---that need better radio efficiency, authentication, and performance than WiFi traditionally offers.

\subsection{Lowering the Barriers}\label{sec:heterogeneous}
\begin{SCfigure}
\centering
\includegraphics[width=0.45\columnwidth]{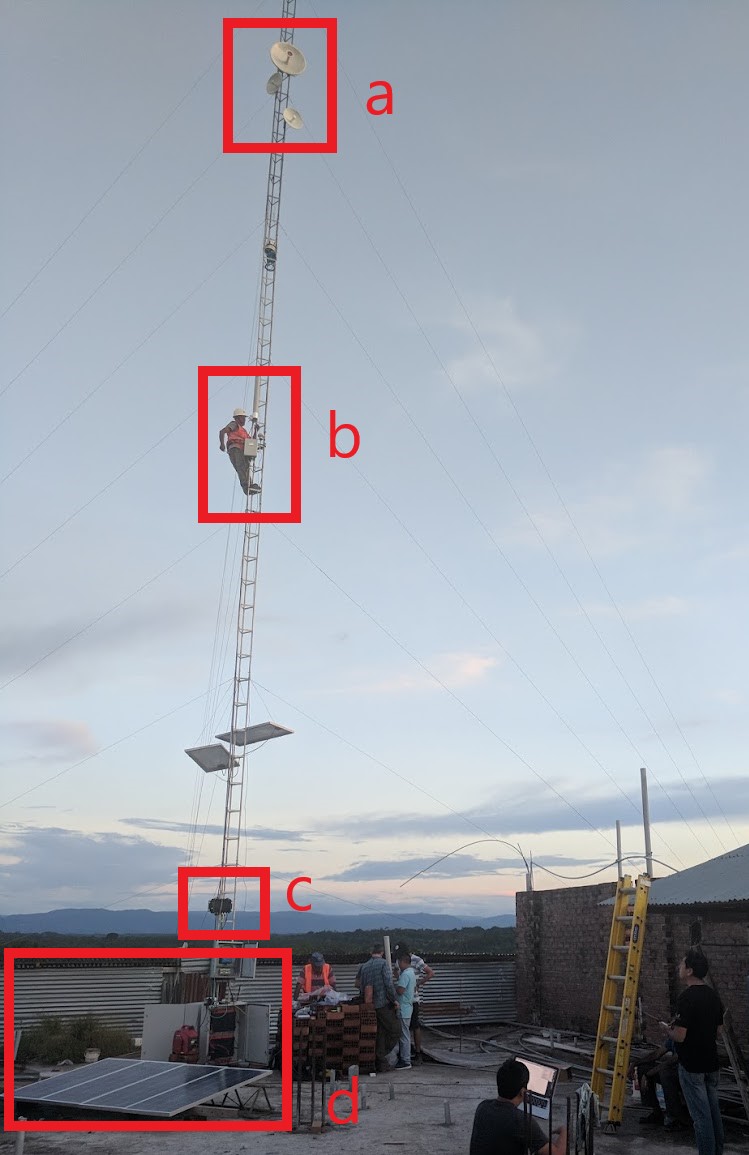}
\caption{An early \name deployment with a small rural ISP in Peru (their first cellular site). Components (top to bottom) include (a) point-to-point wireless backhaul, (b) LTE radio and antenna, (c) ruggedized embedded PC serving as \name AGW, and (d) solar power and battery backup for site.}\label{fig:peru}
\end{SCfigure}

\name aims to lower the barrier to connecting under-served populations via wireless networks. We argue that operators should be able to choose the appropriate access technology for any deployment without then being locked into a core architecture that is compatible only with that access type. 
A single design that supports heterogeneous technologies amortizes the substantial engineering effort for creating software and the costs of training and supporting those who operate the networks.  
Plus, the design enables a single carrier to use multiple radio technologies (e.g., WiFi in shopping malls and cellular elsewhere) on a single core.

Cellular access has high barriers to \emph{entry}.
Network operators deploying cellular access technologies must make large capital investments (CapEx) in infrastructure: whereas a WiFi access point can cost under US\$100, even a low-cost cellular deployment would cost at least 1-2 orders of magnitude more.
Traditionally, cellular core network equipment is designed for large deployments with hundreds of base stations and does not ``scale down'' to small initial deployments at reasonable cost; these networks also have high operational costs (OpEx), relying on highly skilled staff to manage the equipment.
In addition, remote communities may not have affordable access to the high-quality, low-latency backhaul (e.g., fiber) links cellular networks typically rely upon.
Instead, these networks may use satellite or wireless backhaul links with lower reliability and performance.

In contrast, WiFi deployments can 
start small, but present high barriers to \emph{scale}.
WiFi networks do not typically require skilled staff to deploy.
Because WiFi is an inherently best-effort access technology, these networks can leverage any available backhaul, even ad-hoc mesh backhaul using the same physical WiFi radios.
Yet WiFi networks do not typically offer scalable ways to implement network policy or (beyond proprietary and vendor-specific solutions) to manage large networks.
Thus, it is difficult for a WiFi-oriented operator to offer financially sustainable service over a wide area along the lines of large (usually cellular) operators.

Despite the differences between WiFi and cellular, these barriers are \emph{not} fundamental.
The building blocks of network policies are common in each; what is missing is architectural support.
Software-defined networking can help address these gaps by enabling network-wide control over a distributed infrastructure, and adopting ``scale out'' techniques based on commodity components can reduce cost.
In short, adopting and extending successful Internet and cloud approaches to scalability 
and management 
can make it possible to create a wireless access network that is both flexible and affordable.

\section{Magma Architecture}
\label{sec:arch}
\begin{figure}[thb]
\includegraphics[width=\columnwidth]{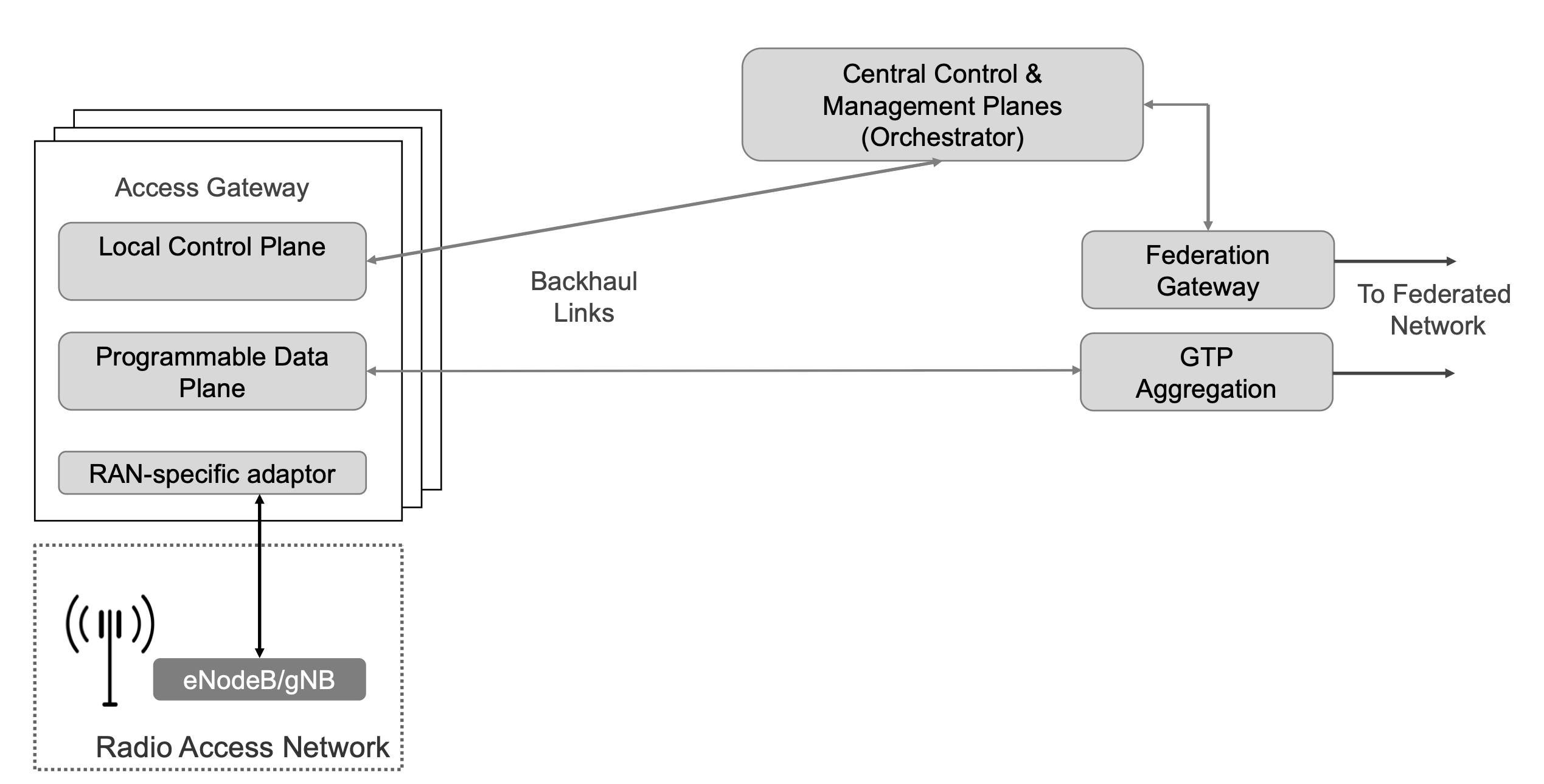}
\caption{Simplified Magma architecture}\label{fig:arch}
\end{figure}

Magma cannot overcome the shortcomings of existing solutions simply by reimplementing a standard, 3GPP-compliant mobile core.  Instead, Magma terminates the radio-specific protocols as early as possible, in \emph{access gateways} (AGWs) connected directly to the radio access network, as shown in Figure~\ref{fig:arch}.  These access gateways are instrumental in handling a variety of radio technologies in a single design.
The Magma architecture goes beyond the traditional RAN/core split of 3GPP to place additional functionality in the access gateway, with a goal of making the packet core more scalable, including scaling down. Notably, Magma adopts the architecture of software-defined networking (SDN) systems, using a hierarchical control-plane design where a local controller in each access gateway interacts
with a centralized \emph{orchestrator}. The orchestrator is the central point of control for the system and maintains authoritative state related to system-wide configuration ({\em config} state). Runtime state, which relates to the activity of user equipment (UEs), is localized to the AGW that serves the appropriate base station for a given UE.

Each AGW is a small fault domain, ensuring that the failure or upgrade of any one component affects relatively few users.
In this way, Magma's architecture is similar to modern cloud systems designed to run on low-cost hardware that is prone to failure~\cite{fox1997cluster}.  Magma adopts other ideas from cloud architectures, including the use of gRPC for communication among components, %
a ``desired state'' model for state synchronization,
and software-based, programmable data plane.
While common in cloud computing deployments, these decisions deviate significantly from the way typical 3GPP networks are designed and managed.

\subsection{Abstracting the Radio Access Technology}\label{sec:abstract}

\begin{figure}
\includegraphics[width=0.95\columnwidth]{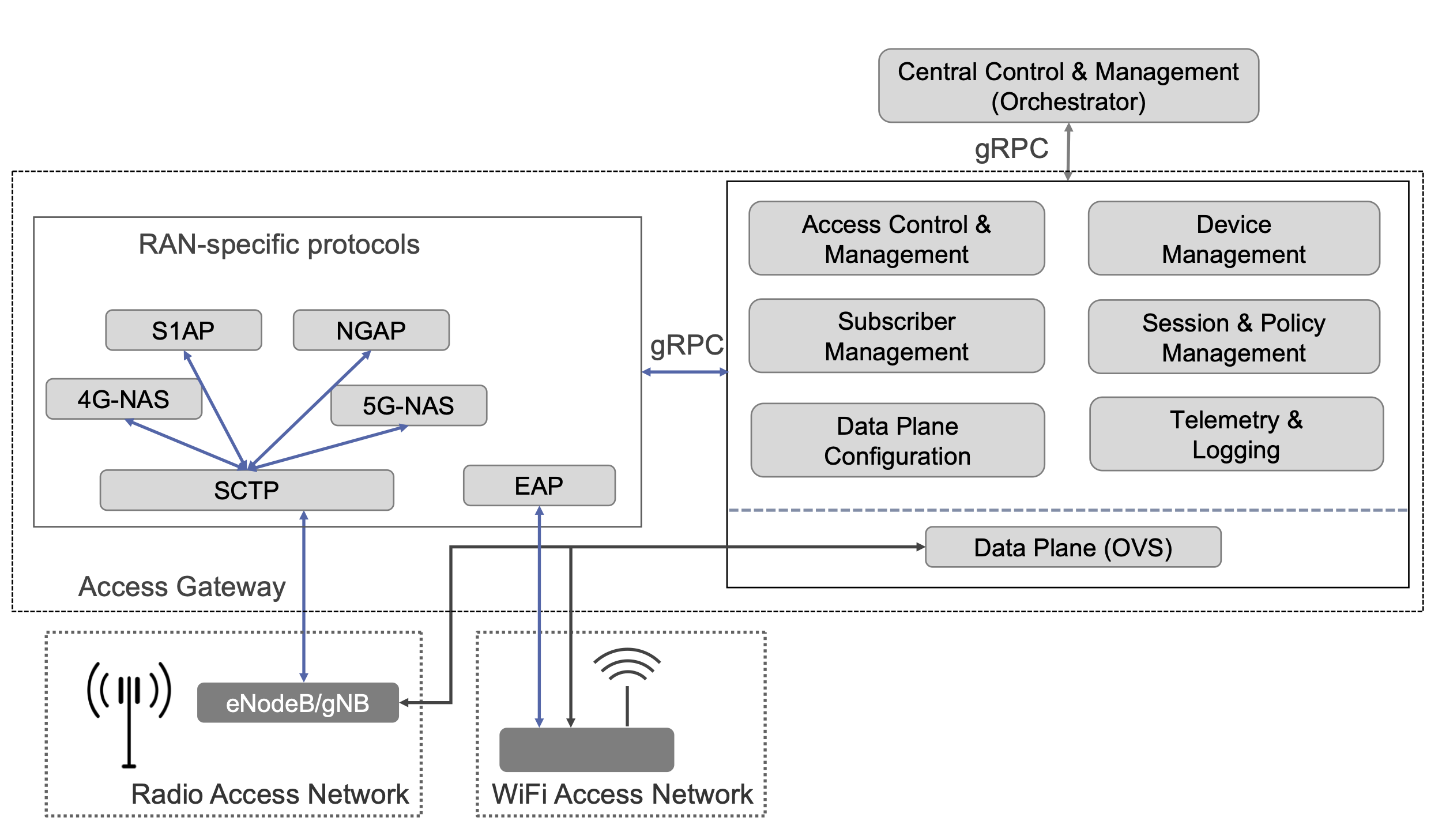}
\caption{Common functions and RAN-specific protocols in the Magma architecture.}\label{fig:common}
    \vspace{-0.2in}
\end{figure}

As Figure~\ref{fig:4g-5g} illustrates, the details of the radio access technology traditionally ``leak'' into the core network. 
To counter this, Magma identifies a core set of functions that the AGW must implement for any radio technology (e.g., finding the appropriate policy for a given subscriber) and provides them in an access-technology-independent way. These functions form the heart of an AGW, as illustrated on the right side of Figure~\ref{fig:common}. 
Control protocols, which are specific to a given radio technology, are terminated {\em early} in technology-specific modules close to the radio. These modules, on the left of the figure, communicate with the generic functions (e.g., subscriber management, access control and management) on the right using messages that are RAN-agnostic. 

Consider the example of ``attaching'' a newly active UE. The UE communicates with a nearby base station over a temporary (unauthenticated) radio link. In traditional 4G implementations, the base station forwards the request to the Mobility Management Entity (MME), which initiates an authentication protocol with the UE. The MME consults a subscriber database, authenticates the UE, creates an entry in a session table, and informs the other components of the parameters needed to serve the UE including: (a) assigning an IP address to the UE and setting the appropriate QoS parameters in the data plane; (b) instructing the base station to establish an encrypted channel to the UE; and (c) giving the UE the symmetric key for the encrypted channel.
At the end of this sequence of events, the UE has an active {\em session} established with the mobile core and is able to send and receive data.

These functions are performed in the Magma AGW in a way that abstracts the details of the radio technology, as illustrated on the right-hand side of Figure~\ref{fig:common}. For example, Magma's subscriber database has the union of all capabilities across the radio access types, even if some fields in a given database row are valid only for some technologies.  QoS policies, for example, are less rich in WiFi than in 4G networks, while 4G policies are in turn less rich than those of 5G. Similarly, UE authentication and session establishment are done in a common way by generic functions that cover 4G, 5G, and WiFi procedures.  The data plane, which is implemented in different devices across 4G, 5G, and WiFi, is implemented in a common, programmable data plane for Magma. 

Table~\ref{tab:mapping} shows how the various components of 4G, 5G, and WiFi are all mapped onto a common set of \name abstractions. The key observation here is that there are a certain set of functions that need to be performed to authenticate users, establish session state, control the data plane, and so on. Magma does all of these in a generic way that is agnostic to the radio technology in use, thus providing an implementation in which the radio-specific details are abstracted from the core and limited to protocol termination close to the radio itself.

Additionally, Magma adds some generic functions that are not part of the 3GPP standards: device management and telemetry. Coupled with the SDN architecture, this simplifies the management 
of a large number of devices spread over a wide geographical area. Rather than logging into a specific device to configure it or check its statistics, Magma provides central management and monitoring from the orchestrator, where it can be leveraged by other systems that consume the northbound API. 
We have found that considering device management and telemetry as first-class responsibilities of \name significantly reduces the operational complexity and cost of operating access networks (Section~\ref{sec:eval:accessparks}).

We do not claim that \name's decomposition of functionality (Figure~\ref{fig:common}) is fundamental, but our operational experience shows that it is useful both from an engineering perspective and for a wide range of use cases (as discussed further in Section~\ref{sec:eval}).
The modularity between components allows \name's internal interfaces to evolve independently of the RAN, aligning with an iterative development approach and in stark contrast to rigid 3GPP interface definitions.
This has enabled the team to perform major changes to AGW functionality, such as adding new features (e.g., 5G support) or refactoring internal services without exposing these changes southbound toward the RAN or northbound toward the \orctr API.

\drop{
\begin{table*}
    \centering
    \begin{tabular}{l|l|l|l} \toprule
        Magma Abstraction & LTE equivalent & 5G equivalent & WiFi equivalent  \\ \hline
        Access Control \& Management & MME & AMF & RADIUS AAA \\
        Subscriber Management & HSS  & UDM and AUSF & RADIUS AAA \\
        Session \& Policy Management & MME and PCRF &  SMF and PCF & RADIUS AAA\\
        Data Plane Configuration & SGW and PGW & SMF & WiFi data plane\\
        Data Plane & SGW and PGW & UPF & WiFi data plane \\
        Device Management & \multicolumn{3}{c}{per-box configuration} \\
        Telemetry and logging & \multicolumn{3}{c}{no equivalent defined}\\
    \end{tabular}
    \caption{The abstractions of Magma mapped onto RAN-specific implementations}
    \label{tab:mapping}
\end{table*}
}

\begingroup
\setlength{\tabcolsep}{4pt} %
\renewcommand{\arraystretch}{1} %
\begin{table}
\scriptsize
    \centering
    \begin{tabular}{l|l|l|l} \toprule
        \textbf{Magma}  & \textbf{LTE}  & \textbf{5G}  & \textbf{WiFi}   \\ \hline
        Access Control/Management & MME & AMF & RADIUS AAA \\
        Subscriber Management & HSS  & UDM/AUSF & RADIUS AAA \\
        Session/Policy Management & MME/PCRF &  SMF/PCF & RADIUS AAA\\
        Data Plane Configuration & SGW/PGW & SMF & WiFi data plane\\
        Data Plane & SGW/PGW & UPF & WiFi data plane \\
        Device Management & \multicolumn{3}{c}{per-box configuration} \\
        Telemetry and logging & \multicolumn{3}{c}{no equivalent defined}\\
    \end{tabular}
    \caption{Magma abstractions vs. RAN-specific versions}
    \label{tab:mapping}
    \vspace{-0.1in}
\end{table}
\endgroup

All communication between the RAN-specific modules on the left of Figure~\ref{fig:common} and the generic functions on the right use gRPC~\cite{grpc}, an open-source Remote Procedure Call (RPC) framework, as does all long-distance communication (e.g., from the AGW to the \orctr).
Although this is a typical approach for building modern distributed systems, it differs substantially from the protocols defined for communication among 3GPP components, which leak endpoint (e.g., UE and MME) consistency requirements into a network-level protocol.
By running over HTTP, gRPC inherits the resilience to loss and delay of TCP/IP, which is absent from some 3GPP protocols designed for more benign, 
controlled environments (e.g., leased lines).
A concrete example is GTP (GPRS Tunneling Protocol), which is sensitive to loss and latency to the point that it struggles to operate over lower quality or congested backhaul links, such as satellite or shared microwave links.
Thus, adopting gRPC allows \name more latitude to implement alternative consistency models without breaking UE state machines in a wider range of backhaul network conditions.
In practice, this tolerance helps mitigate poor error handling on devices: while UEs should reconnect after experiencing a 3GPP protocol-level failure, we find that UEs with low-end baseband processors do not do so reliably.
When a UE fails to reconnect, the failure manifests as a confusing lack of coverage to people using these devices, and the failure typically only resolves after power cycling the UE. Since \name terminates GTP locally in the AGW without traversing the backhaul link, a UE never sees a dropped GTP connection and does not have to handle the error.

While agnostic to the radio technology, Magma necessarily makes practical choices about the order of feature development. Many early Magma deployments used LTE, so we have prioritized support for LTE features, with 5G support coming later. A good example is the support of QoS policies.  Simple policies to impose rate limits and usage caps, as outlined in Section~\ref{sec:heterogeneous}, are supported today for both LTE and 5G. More complex policies could be expressed, particularly in 5G, but full support for richer policies is currently under development.

\subsection{Hierarchical SDN Control Plane}\label{sec:hierarchy}
Magma adopts software-defined networking (SDN) to reduce operational complexity and minimize reliance on skilled staff.  Rather than configuring a distributed collection of devices, providers specify network-wide policies at the orchestrator.  The orchestrator provides a central point of control and exposes a northbound API for integration with other systems (e.g., for metrics, alerting, and monitoring). However, running the entire control plane in a central controller would impose limits on the scalability of the system.  Hence, practical SDN systems like Network Virtualization Platform (NVP)~\cite{nsdi14-nvp} and Open Virtual Network (OVN)~\cite{OVN} adopted a hierarchical control plane, and this is the model used by Magma.

In a hierarchical control-plane design, we identify those elements of the control plane that have network-wide scope; these are candidates for the central controller. For example, to add a new subscriber to the network, the long-lived information about the subscriber is network-wide information that is created and maintained by the central controller. Conversely, much of the runtime state associated with a UE can be localized to a single AGW. For example, upon becoming active, a UE is associated with a single AGW. The UE's session state can be created and managed by the local control plane of that AGW. Thus, much of the control plane is able to scale out with increasing numbers of base stations and subscribers, rather than increasing the processing in the central controller.

This division between central and local control planes roughly corresponds to the timescale of changes to the control-plane state. The addition of a new subscriber happens on configuration timescales, and that state is managed centrally. The creation of session state---when a UE becomes active and attaches to an AGW---happens more frequently. This runtime state is handled by the local controller on an AGW. 

As with any SDN architecture, we must consider ``headless'' operation, i.e., the situation where a data plane node is disconnected from the central control plane. In a traditional SDN approach, the goal is to ensure that the data plane continues to operate without the control plane, even as updates to the data plane may be impossible while the control plane is disconnected. With a hierarchical control plane, many local operations are still possible even while the central controller is unreachable. For example, an AGW can still establish a session for a UE that attaches to a base station, because the local control plane has enough information (e.g., cached subscriber profiles) to process the session establishment. Conversely, network-wide actions like the addition of users or changes to user policies must wait until the central control plane becomes available again.  Magma makes trade-offs for availability versus consistency as the CAP theorem \cite{CAP,brewer2012cap} implies. It is generally possible for state stored in an AGW to be stale during times of disconnection, which might, for example, allow a UE to temporarily consume resources beyond its quota. 

This design helps to achieve the scaling goals of Magma, in allowing both a small minimum footprint (scaling down) as well as scaling up. A minimal Magma deployment would be a single AGW and an orchestrator. The orchestrator is typically three virtual machine instances in a cloud, while the AGW itself is a small (4-core) x86 commodity server. This is dramatically less hardware than a conventional cellular packet core.
Scaling up is essentially a matter of adding more AGWs, which increases the number of base stations and UEs, without much increase in the load on the orchestrator. We discuss our experiences in scaling up in Section~\ref{sec:deploy}.

The decision to place local control-plane functions on the AGWs, while beneficial for scalability, does introduce trade-offs. In particular, it complicates the picture for some features that require coordination among AGWs. Notably, while \name supports mobility across radios served by a common AGW, seamless mobility {\em between} AGWs would require communicating some control-plane state from one AGW to another during hand-offs. While many use cases can be supported without this feature, we expect to add it in the future.

\subsection{Fault Tolerance Via Small Fault Domains}
\label{sec:fault-tolerance}
The desire to build a low-cost solution for Magma has a significant effect on how the architecture approaches fault tolerance. Low-cost hardware is prone to failure, and so Magma adopts the view common to most modern cloud systems: it is expected that individual components will fail.  A failure of a component must affect as few users as possible (i.e., fault domains must be small) and must not affect other components.  This approach also has a positive impact on operations such as software upgrades, as it is possible to upgrade small components independently without taking down the whole system. This is in stark contrast to traditional 3GPP implementations.

The SDN-like architecture of Magma localizes state more fully than a typical 3GPP implementation. In a standard implementation, the runtime state of a UE is spread among several large components (e.g., the PGW, SGW, and MME in the LTE case).  In contrast, Magma localizes the runtime state of a UE to a single AGW. This simplifies failure handling. The runtime state stored in an AGW is checkpointed regularly and may be copied to a backup instance of the AGW running as a cloud service. When an AGW fails, the backup cloud instance is brought into service, and can manage connections for the affected set of UEs until the primary AGW is restarted.
 As noted above, an AGW may continue to establish sessions to UEs even when disconnected from the orchestrator. The state synchronization approach described in Section~\ref{sec:desired-state}  mitigates the long-term effects of such failures. 

While it is common for a traditional cellular packet core to serve millions of subscribers,  Magma  distributes much of the functionality  to a large number of Access Gateways. Each AGW is thus a fault domain that holds state for a relatively small number of UEs served by a small number (typically less than ten) of base stations. The failure of a single AGW would impact the set of UEs currently served by the attached base stations, but has no impact on the rest of the network or its customers. This contrasts with the much larger fault domains typical of a standard mobile core implementation.

\subsection{State Synchronization}
\label{sec:desired-state}
State in a mobile core needs to be communicated among components. Generally, one component is the authoritative owner of some piece of state, and it needs to synchronize state with another component. %
In \name, state can take one of three forms, for which \name makes different guarantees.

The first is \emph{runtime state} associated with a UE and its network activity.
Backwards compatibility with existing user devices and RAN equipment requires \name to implement standards-defined state machines to support operations like connecting to the network; modifications to runtime state can occur due to events in the UE itself, the RAN equipment, or  Magma's ``core'' network elements.
Importantly, runtime state within \name is encapsulated within the AGW, which as discussed in Section~\ref{sec:fault-tolerance} is the failure domain for \name, and we assume a crash-recovery failure model for AGWs.\footnote{We generally assume the same for individual AGW software components; per-process state is held externally for most critical services.}
Further, most runtime state is both ephemeral and recoverable in the event of failure: a UE can simply reconnect.

The second is the \emph{configuration state},
associated with the configuration of a \name network element, such as an AGW.
This is only ever written by the \orctr and pushed asynchronously to the AGW. 
Examples include classes of network policy to be applied to classes of user or radio configuration to be applied by an AGW to connected RAN equipment.
AGWs recover configuration state after a crash, and the source of truth for configuration state is stored durably in the \orctr (Postgres); we only permit modification to configuration state through the \orctr.
Configuration state generally changes on human timescales (i.e., minutes or hours).

Finally, \name also manages \emph{metrics state}, telemetry from Magma elements.
This operational data, while useful, is captured on a best-effort basis.

Like many cloud-native systems, Magma adopts a ``desired state'' model for runtime and configuration state. By this we mean that to communicate a required state change (e.g., the addition of a new session in the data plane), the desired end state is set via an API. This contrasts with a ``CRUD (Create, Read, Update, Delete)'' interface, which is common in 3GPP specifications. Magma replaces the CRUD model with the desired state model to simplify reasoning about changes across elements of the system in the case of partial failures. This is a common case in challenged environments, where portions of the end-to-end system (e.g., backhaul) are far less reliable than others (e.g., the link between the UE and the RAN). This is best explained via a simple example.

Consider the case of establishing data-plane state in an AGW for a set of active sessions. Suppose there are two active sessions, X and Y. Then a third UE becomes active and a session Z needs to be established. In the CRUD model, the control plane would instruct the data plane ``add session Z''. The desired state model, by contrast, communicates the entire new state: ``the set of sessions is now X, Y, Z''. 
The CRUD model is brittle in the face of failures. If a message is lost, or a component is temporarily unable to receive updates, the receiver falls out of sync with the sender. So it is possible that the control plane believes that sessions X, Y and Z have been established, while the data plane only has state for X and Y. By sending the entire desired state, we ensure that the receiver comes back into sync with the sender once it is able to receive messages again. 

This approach is hardly a novel idea in the cloud-native world, but it differs from typical 3GPP systems. It allows Magma to tolerate occasional communication failures (caused by poor quality backhaul, for example) or component outages due to software restarts, hardware failures, etc. Limiting the scope of 3GPP protocols to the very edge of the network gave us the flexibility to rethink state synchronization to improve fault tolerance (in addition to other benefits noted above). 

We close by considering how \name manages state for one particularly salient policy: billing users based on data volume, and the possibility of double-spending.
Volume-based billing policies are typically implemented using a third-party online charging system (OCS) that integrates with both the network operator's existing business support systems (BSS) as well as \name.
In this arrangement, \emph{billing and charging} are handled by the OCS, while \name handles \emph{metering and accounting}.
The OCS tracks a user's account balance (e.g., in US\$) and then authorizes small quotas of data (e.g., 1MB) to the user via \name; when the user nears completion of their quota, \name requests another quota on the user's behalf from the OCS, which makes the decision on whether to grant or deny the request.
Whether or not a user has been allocated a quota is \emph{configuration} state from \name's perspective, while the amount remaining in a user's current quota is \emph{runtime} state.
Thus, while it is possible for a malicious user to double-spend by moving between AGWs strategically, the maximum amount of double-spend permitted is capped as a business decision by the quota size.
Operators for whom this is a particular concern could also adopt techniques for volume-based accounting in a distributed context~\cite{hasan2019ccm}.

\subsection{Software Data-Plane Implementation}
\label{sec:data-plane}
The data plane is responsible for
(i)
recognizing the flows for active sessions (traffic to and from active UEs);
(ii) collecting statistics for those flows;
(iii) adding and removing tunnel headers; and
(iv) enforcing policies such as rate limits per subscriber.
Magma's data plane is implemented using Open vSwitch (OVS)~\cite{ovs}. OVS provides a programmable data plane that is controlled by OpenFlow~\cite{openflow}. 
While OpenFlow and OVS are convenient implementation choices, they are not fundamental to the architecture. Other options may be used in the future. The important points are that the data plane is highly programmable and implemented entirely in software.

The software implementation of the data plane enables Magma to operate on commodity hardware. While throughput, latency, and jitter of the data plane are important for cellular networks, we have found OVS to offer entirely adequate performance. OVS performance has been well studied and optimized for many years \cite{ovs}. In Section~\ref{sec:eval} we evaluate the performance of OVS in the Magma context. It is worth noting that other aspects of the system such as backhaul and RAN capacity are likely to have a larger performance impact overall than the data plane within the access gateway.

The ``data plane configuration'' box in Figure~\ref{fig:common} generates the commands necessary to program the data plane with a set of rules to handle the flows of current sessions. Currently, those commands are OpenFlow commands. If OVS were replaced with a different  forwarding engine, only the ``data plane configuration'' component would be affected.

\subsection{Federation With Other Networks}
\label{sec:federate}

To this point we have described standalone deployment of \name, but it can be deployed in one of three modes:
\begin{CompactItemize}
\item \emph{Standalone}: \name supports an independent network, with all 3GPP control and user plane traffic terminated in the AGW.
\item \emph{Local breakout roaming}: \name{}  {\em federates} with an existing cellular network, with control-plane traffic terminated externally but user-plane traffic still handled by the AGW and routed directly to or from the Internet.
\item \emph{Home roaming}: \name federates with an existing cellular network, with both control and user-plane traffic terminated in an external network.
\end{CompactItemize}

Much as the AGW terminates access-specific protocols from the radio network, \name introduces additional elements to terminate access-specific protocols with an external core network, using a component referred to as a \emph{Federation Gateway} (FeG). 
The FeG implements 3GPP-defined interfaces to support ``home roaming'' as well as ``local breakout roaming''. 
The latter is made possible in \name by the fact that rich policy enforcement is provided in the AGW. As an example, an AGW can obtain the policy to apply to a UE by querying the subscriber data base in the federated network, then enforce that policy in the AGW. 
Signalling traffic between UEs and the MNO core is handled by the FeG service in the \orctr\footnote{This is necessary to coordinate low-level network state between the UE and the MNO's traditional core, such as GTP bearer identifiers.}.
User data-plane traffic is tunneled to an analogous component, the GTP Aggregator (GTP-A) which in turn connects to the MNO's existing P-GW.

Unlike the AGW, the FeG and GTP-A are centralized, on-path devices.
This serves a practical purpose:  traditional MNOs prefer a single point of interconnection between their sensitive core network and ``extension'' networks~\cite{hasan2019ccm}. This has scaling implications as discussed in Section~\ref{sec:eval_fed}.

\section{Evaluation}
\label{sec:eval}
\name makes a number of fundamental design choices that differ from traditional core network software to improve flexibility and scalability, while supporting rich network policies.
The aim is to support \emph{practical} cellular access deployments.
To evaluate \name, we  first consider 
system performance in an emulated environment, and  then discuss a large-scale commercial \name deployment.

\subsection{Supporting Typical Deployments}

\textbf{Emulation Testbed}
Although evaluating \name's performance in a real deployment is possible at small scale, evaluating scenarios with hundreds of UEs and RAN elements is impractical. Further, extracting data from commercial deployments  is challenging due to privacy and commercial considerations. %
Thus, we instead evaluate \name using a commercial 
emulation system, Spirent Landslide~\cite{landslide}, which allows us to emulate arbitrary configurations of virtual UEs and RAN elements in a replicable fashion.

For our evaluation, we deployed the most recent stable release of \name, v1.6.1. We deployed the \orctr on a cluster of AWS EC2 instances and two AGWs in our lab.
The first AGW was a bare-metal AGW on an Intel J3160 quad-core 1.6GHz CPU with 8GB of RAM and four Intel I210 1Gbps NICs.
The second was a virtual AGW running with Intel Xeon 6126 2.60GHz, 8GB of RAM, and 2x10G Mellanox ConnectX-3 NICs; we assigned a variable number of vCPUs to the virtual AGW as defined in our experiments below.
Both the bare-metal and virtual AGWs were connected directly to the Landslide emulator as well as to the Internet via 1Gbps and 10Gbps links, respectively.
We also verified that memory was not a bottleneck for the AGW during our experiments and that all machines in the \orctr deployment were running well under capacity.
Finally, the emulated SIM cards for the emulated UEs were pre-provisioned into the \orctr and AGW in advance of all experiments, as is typical for network operator deployments of Magma.

Unlike traditional core networks, \name's AGW is co-located with RAN equipment (for example, at a tower site), and the unit of scaling for \name is the AGW itself: as operators grow their network, they add both additional RAN capacity (i.e., radio equipment) but also additional ``core'' capacity (i.e., AGW instances).
Since the AGW is an on-path device for all traffic associated with the cell site, the AGW should be provisioned such that site is limited by the capacity of the RAN as the site, rather than the AGW.
This is a notable observation that, in part, motivates Magma's design: when co-locating core network functionality with RAN elements, the \emph{RAN is the bottleneck} for performance on a per-site basis.

The recommended (and typical) deployment scenario has roughly one AGW per ``cell site", which in practice consists of 1-3 eNodeBs in the case of an LTE network.
A typical eNodeB (such as those described in Table~\ref{t:site-cost} or depicted in Figure~\ref{fig:peru}) can support at most 96 simultaneously active users\footnote{More users may be attached but not actively transmitting data.} and radio channels of at most 20MHz; this channel capacity, in turn, corresponds to a peak aggregate throughput of 126Mbps~\cite{3gpp-ts36.213} under ideal conditions, for a typical cell site maximum capacity of 378Mbps.
We note that the additional cost of an AGW is modest in comparison to the cost of a cell site, similar to the site cost breakdown observed in related work~\cite{hasan2019ccm}.
Although LTE site costs can vary widely and are, in our experience, dominated by non-networking costs such as land, power, and tower (also known as ``passive infrastructure''), a representative deployment could consist of the hardware in Table~\ref{t:site-cost}; AGW cost represents less than 3\% of the cost of active equipment for the site. Power costs can be especially significant in ``off-grid'' locations, but these are largely driven by the power needs of the radio equipment and hence not greatly influenced by the mobile core implementation. Note the use of solar and battery power in Figure~\ref{fig:peru}.

\begin{figure}
\centering
\includegraphics[width=\columnwidth]{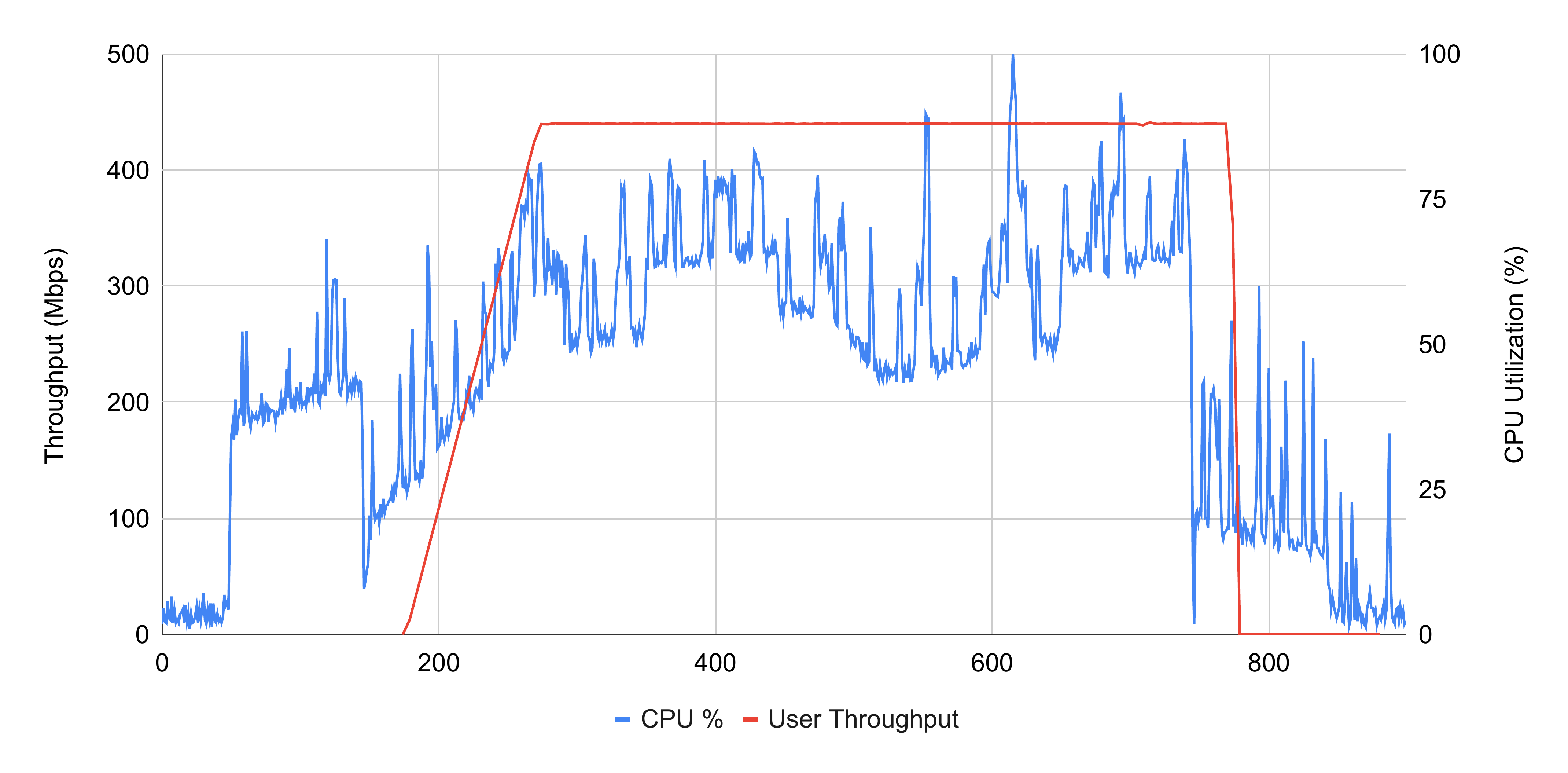}
\vspace{-.4in}
\caption{AGW CPU utilization under maximum ``typical'' workload for a cell site. 
Aggregate throughput is limited by radio capacity, not the AGW.}
\label{fig:exp1a_cpu}
\end{figure}

\name must be able to support this type of workload.
We evaluate this by emulating the peak load of a the cell site described above: a total of 288 UEs connect (or ``attach'') to the network for the first time at a rate of 3UE/sec, and each then performs a short HTTP download at a rate of 1.5Mbps, for an aggregate total offered load of 432Mbps.
Figure~\ref{fig:exp1a_cpu} demonstrates our results, focusing on the total CPU utilization as well as achieved throughput of the AGW.
At a high level, the AGW accepts attach requests from all new users over the course of approximately 1.5 minutes, after which the AGW enters a steady state for the duration that UEs are making HTTP requests.
In this experiment, average sustained UE throughput reaches the expected throughput of 432Mbps throughout the duration of the experiment, indicating performance is limited by the RAN, rather than \name's AGW, as expected.

\begingroup
\setlength{\tabcolsep}{4pt} %
\renewcommand{\arraystretch}{1} %
\begin{table}[t]
\scriptsize
    \begin{center}
        \ra{1.06}
        \begin{tabular}{@{}p{1.5cm}rrrp{3cm}@{}} \toprule
            \textbf{Item} & \textbf{Unit Cost} & \textbf{Qty} & \textbf{Total} & \textbf{Notes} \\ \hline
            LTE eNodeB & US\$4,000 & 3 & US\$12,000 & Baicells Nova 223: 1W, 3.5GHz, 96 user, 2x2 MIMO. \\
            AGW & US\$450 & 1& US\$450 & Same as used in experiments.\\
            Accessories & US\$450 & 3 & US\$1,350 & 18dBi sector antenna, RF cables, connectors, grounding. \\
            \hline
            \multicolumn{2}{l}{\textbf{RAN CapEx (per site)}} & & US\$18,760 &  \\ 
            \bottomrule
        \end{tabular}
    \end{center}
    \vspace{-12pt}
\caption{Cost breakdown of active RAN equipment for a typical \name deployment. Excludes site-specific passive infrastructure and backhaul costs.}
    \label{t:site-cost}
    \vspace{-0.1in}
\end{table}
\endgroup

We acknowledge that other RAN configurations can exist (including vRAN/cRAN arrangements) where many RAN elements are effectively ``co-located'' to a single point within the operator’s network. 
Magma can  be used effectively in these deployments, with one (or more) AGWs  allocated to support this range of RAN equipment. 
However, no deployments at scale of Magma to date have used that configuration (to our knowledge), and Magma can be deployed on any general-purpose compute (e.g., VM or container) alongside this RAN infrastructure to support it.
Similarly, a radio vendor could integrate an AGW into the same physical enclosure as a traditional eNodeB for a combined RAN and AGW element.

\subsection{Control and User Plane Separation}
Different usage patterns of a network stress user plane or control plane elements of the network core: a common example of the former would be human users accessing video content while the latter would be an IoT workload consisting of large numbers of devices that only exchange occasional small messages.
This presents a major dimensioning challenge in traditional cellular core networks and motivates efforts to separate control and data plane elements so operators can scale them independently (statically or dynamically); this is known as ``control/user plane separation'' (CUPS) in LTE and 5G.

\begin{figure}
\centering
\includegraphics[width=\columnwidth]{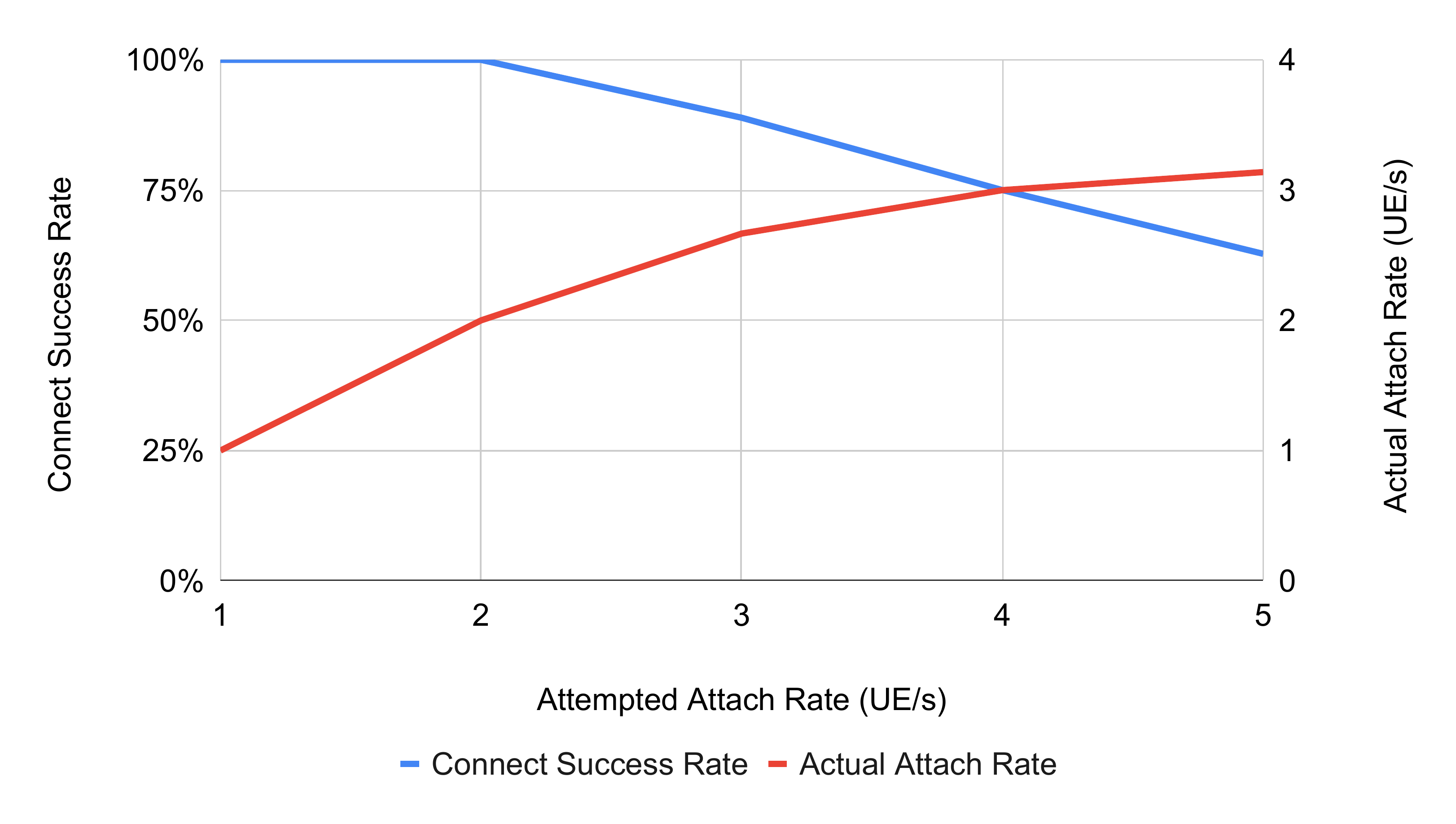}
\vspace{-.4in}
\caption{
Maximum supported attach rates are limited by the AGW (specifically, the MME component). Results depict a physical AGW.
}
\label{fig:exp1b}
\vspace{-.2in}
\end{figure}

\name's distributed design naturally facilitates a CUPS architecture.
By default, every AGW implements a data plane at the network edge, and all control plane functions are implemented as user-space processes at the AGW, with configuration state managed by the \orctr.

From Figure~\ref{fig:exp1a_cpu}, we observe that the AGW operates in two distinct and characteristic domains.
At the start of our experiment, while UEs are attaching to the network, the AGW's CPU workload is dominated by the \emph{control plane} workload associated with handling attach requests, including performing cryptographic operations necessary to authenticate users as well as setting up per-user, per-session state in the data plane and control plane to implement the desired policy for each UE; in our experience, this is the most computationally-intensive control plane procedure.
After UEs attach, the CPU workload is dominated by \emph{user plane} workload associated with forwarding UE traffic.

Figure~\ref{fig:exp1b} illustrates how our bare-metal AGW copes with a ``worst case'' control plane workload, a surge of new UEs attaching then saturating the data plane.
We define the \emph{connection success rate} (CSR) to be the number of connection attempts that succeed over the total number of connection attempts made, for each five second bin during the experiment.
We observe that above 2UE/s, the bare-metal AGW is unable to service all connection attempts, with the connection success rate (CSR) falling linearly beyond this point.
On a per-AGW basis, \name's control-plane performance is relatively limited; improving this is an active area of engineering effort.
Attach rate is a function of hardware as well: a 4 vCPU instance of our virtual AGW supports 16 attaches per second, which would saturate the RAN capacity of the ``typical'' site described above in 18 seconds.

\begin{figure}
\centering
\includegraphics[width=0.95\columnwidth]{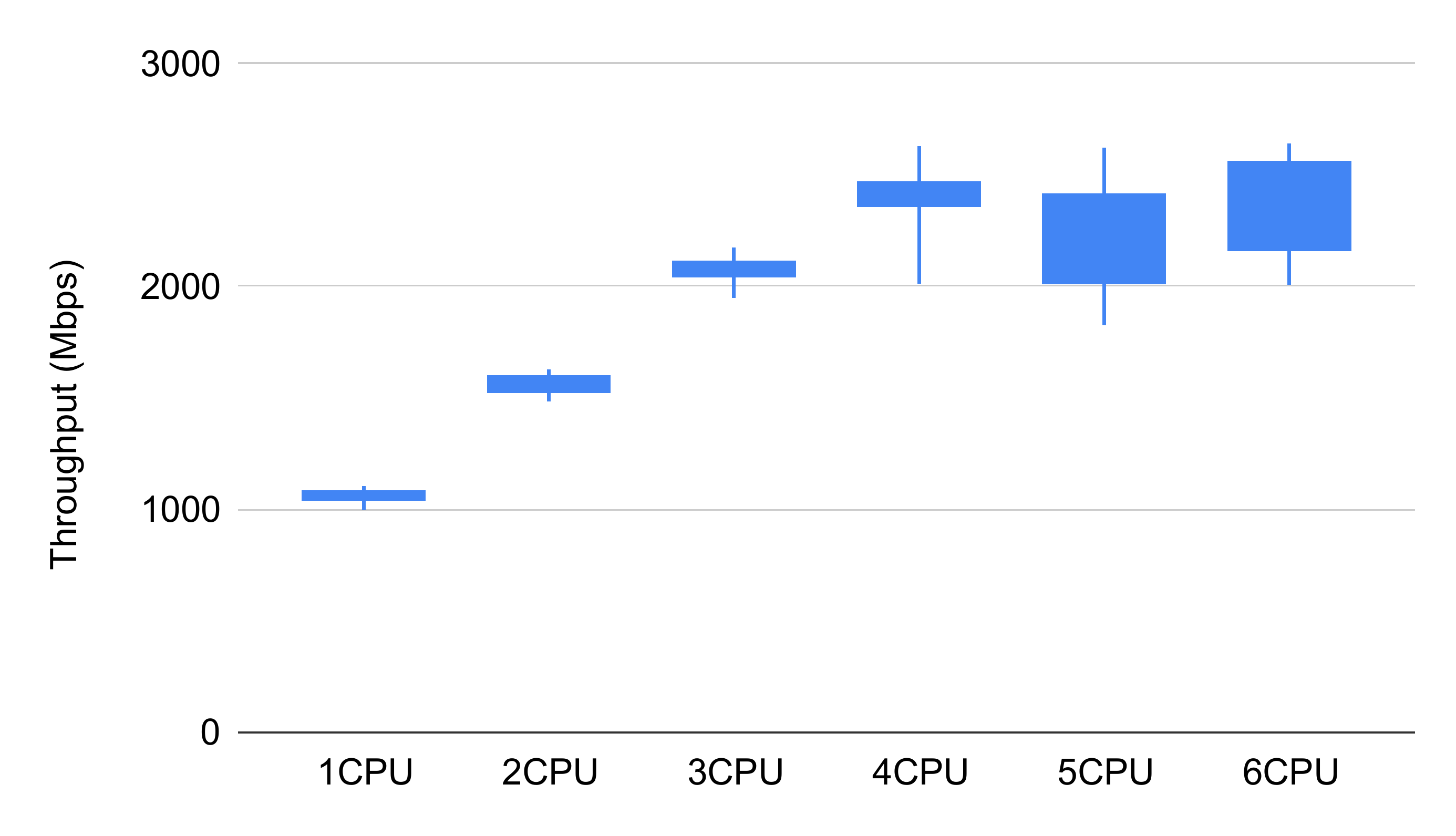}
\vspace{-.2in}
\caption{
Steady state throughput vs. CPUs allocated to user plane. Note our traffic generator was unable to saturate the virtual AGW's user plane in the 5CPU case and above.
}
\label{fig:1c_cpu}
\vspace{-.3in}
\end{figure}

\begin{figure}
\centering
\includegraphics[width=0.9\columnwidth]{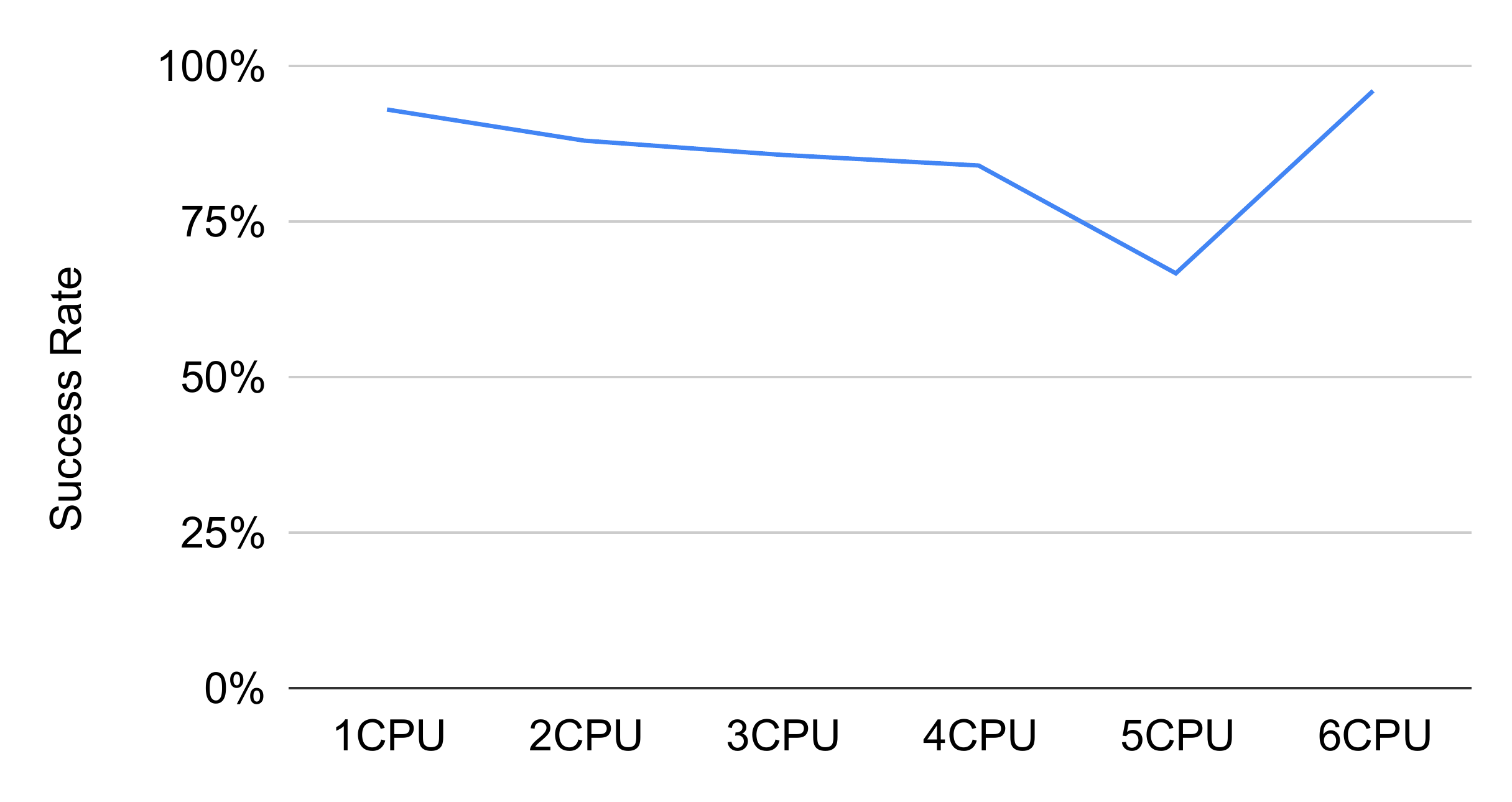}
\vspace{-.2in}
\caption{
Median connection success rate vs CPUs allocated to user plane.
}
\label{fig:1c_csr}
\vspace{-.1in}
\end{figure}

Lastly, we consider per-AGW allocation of resources to the control and user plane.
To do this, we statically limit the number of cores available to the user plane and evaluate steady-state throughput and median connection success rate.
These results are shown in Figures~\ref{fig:1c_cpu} and~\ref{fig:1c_csr}; note that these experiments use the VM AGW, and as such the absolute throughput numbers are not comparable with earlier experiments.
We observe that increasing the cores available to the user plane improves steady-state throughput at the cost of decreased connection success rate (i.e., control-plane performance), but allowing the kernel scheduler to allocate resources flexibly between user plane and control plane tasks provides both high throughput and good connection success rates.
We note that we expect raw user-plane performance to increase beyond what is shown here; the commercial test equipment we used was unable to generate more than 2.5Gbps aggregate load.

Taken together, these emulation results demonstrate that \name can handle typical workloads using low-cost commodity hardware.
For more intensive workloads, \name's control and user plane capacity scales with additional hardware.
We finally note that these results provide an upper-bound on the performance of a \emph{single} \name AGW; the \emph{network} capacity of a Magma network scales linearly with AGWs.

\subsection{Deployment}\label{sec:deploy}
We now turn to large commercial deployments of \name.
We first note that \name is an open-source project governed by the Linux Foundation, and as such the core development team (including the authors of this paper) do not directly operate any production deployments; as such, we draw our examples from partners within the project's ecosystem.

\textbf{Magma adoption.} To understand how \name is used in practice, we interviewed two people working in product management and marketing for the \name open-source project; in their roles, they speak regularly with  operators as well as other commercial entities within the \name ecosystem.
Based on our discussion, as of February 2022, twenty commercial networks were operating using Magma across eight countries in Africa, Asia, North America, and South America.
These networks support a range of access modalities and policies.
For example, \name has been used in networks providing backhaul for WiFi hotspots, fixed wireless broadband to homes and businesses, ``carrier'' WiFi to extend a traditional mobile operator's service to indoor WiFi, and traditional mobile broadband service.
Today, \name has approximately 100 active committers to its codebase.

\textbf{Magma deployments.} To demonstrate how \name is used, we worked with one of the largest commercial entities, \ffi, that provides support to  operators deploying Magma.
\ffi provided data to characterize two significant deployments they help operate.
This data was provided to the authors in de-identified form, and only operational data (not user data) was used in our analysis.%

\subsubsection{Fixed Wireless Hotspots}
\label{sec:eval:accessparks}
\begin{figure}[bh]
\includegraphics[width=0.9\columnwidth]{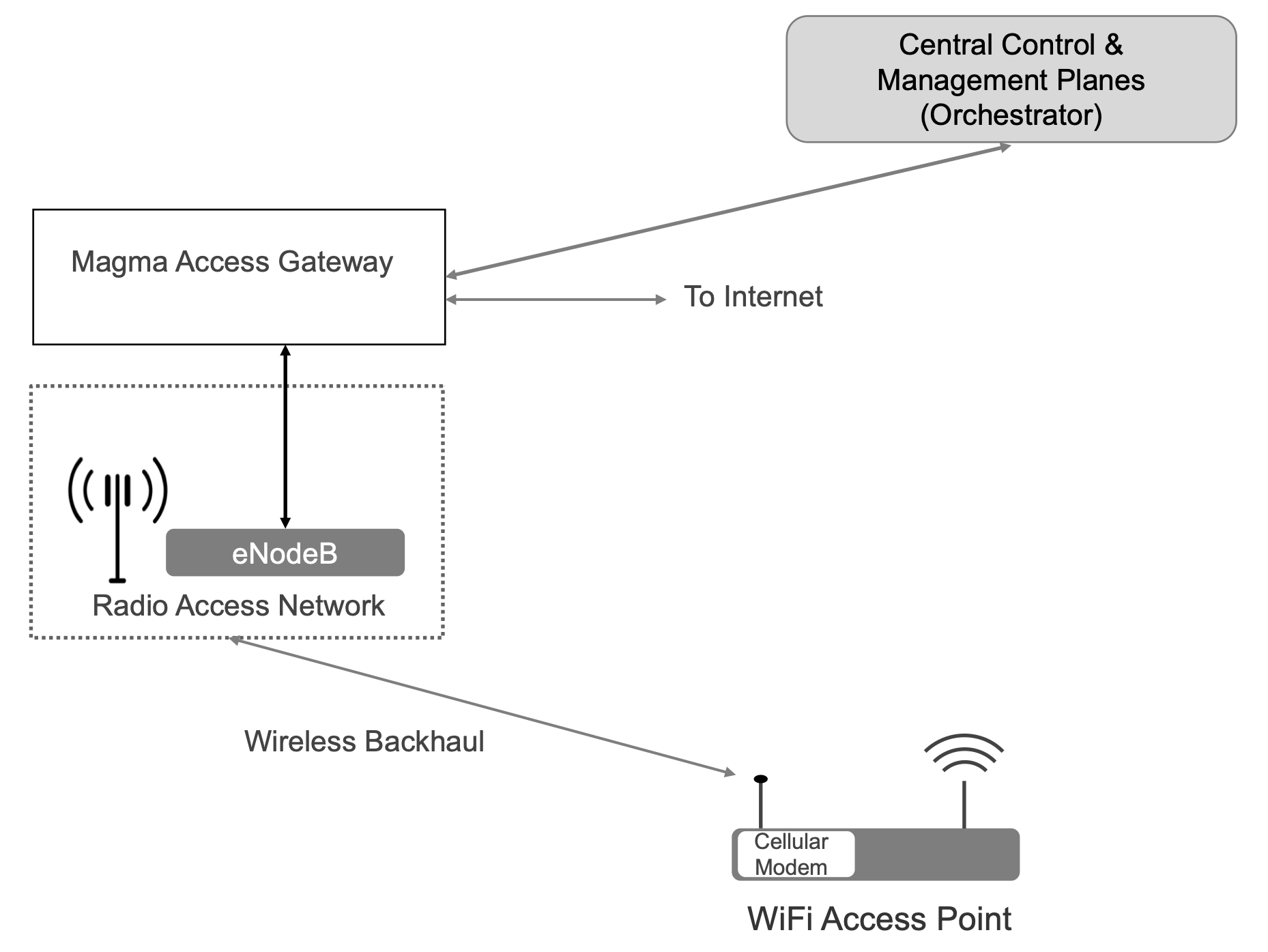}
\caption{Wireless backhaul to WiFi hotspots provided by Magma. This is the network architecture used by \aparks in their deployment: end users connect to WiFi access points via standard mechanisms, and traffic is backhauled from the hotspot via a co-located cellular modem to the LTE RAN supported by Magma. Note that nothing in this design precludes an end user from directly connecting to the LTE network, if appropriately configured and allowed to do so by the network operator.}\label{fig:hotspot}
\vspace{-0.2in}
\end{figure}

One of \ffi's first commercial  deployments was \aparks~\cite{accessparks}, %
a US-based operator that provides public WiFi hotspot networks in large outdoor areas; their deployment locations require multiple WiFi access points (APs) to provide consistent service.
With the availability of CBRS spectrum, \aparks sought to use LTE to provide backhaul to their WiFi hotspots in some of their larger deployments.
End users connect to \aparks's WiFi access points via traditional WiFi mechanisms and an existing captive portal system, and the UEs in the \name network are fixed wireless modems that connect the WiFi APs to the Internet via \name. The setup is illustrated in Figure~\ref{fig:hotspot}.

\begin{figure}
\centering
\includegraphics[width=\columnwidth]{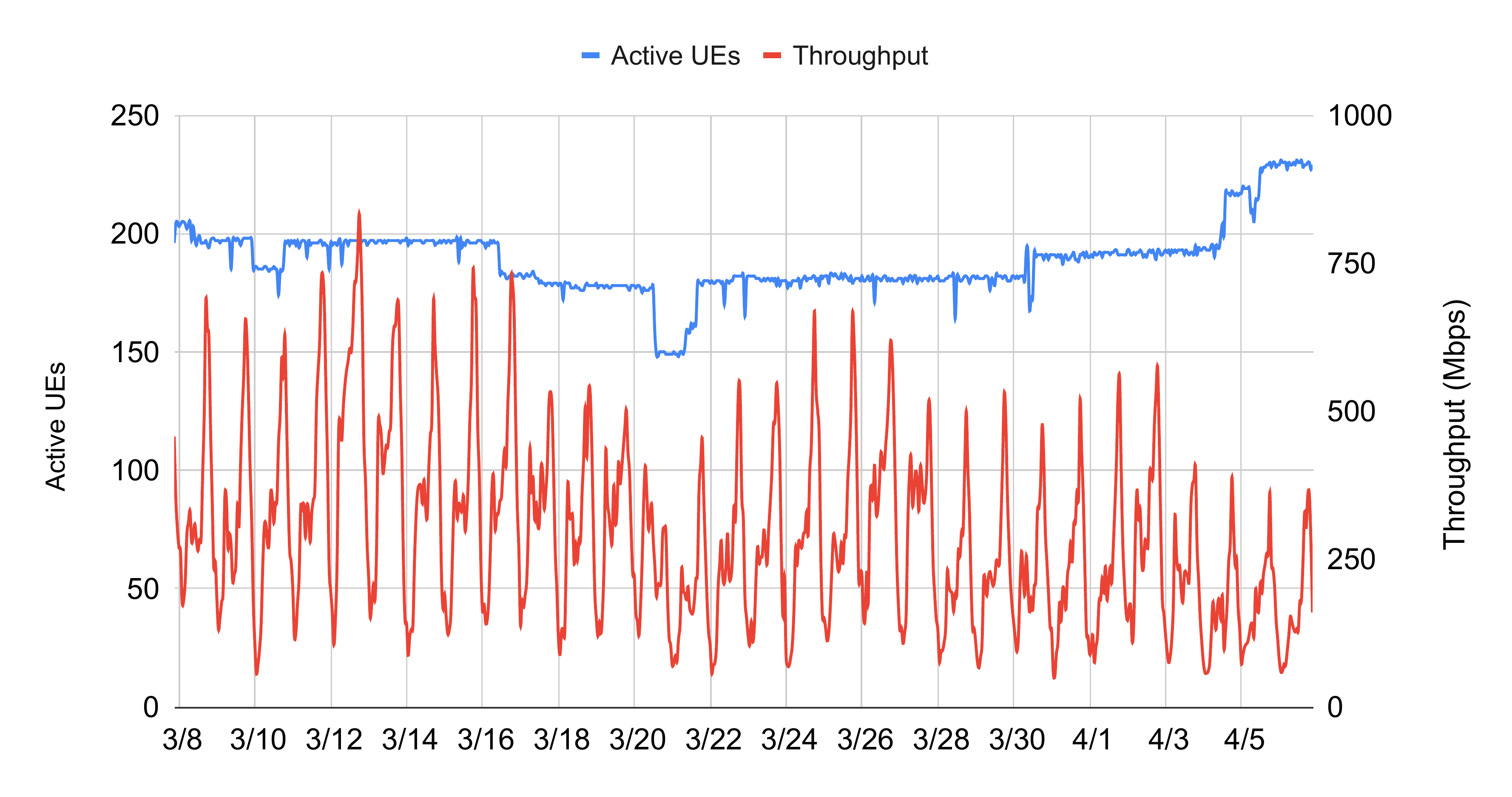}
\vspace{-.3in}
\caption{Per-hour \aparks usage during Mar-Apr 2022.}
\label{fig:ap_usage}
\vspace{-.1in}
\end{figure}

\aparks's deployment began in December 2020 with a ten site pilot to evaluate \name.
Today, the network consists of fourteen sites providing backhaul to over 200 access points, with plans to continue expanding.
Figure~\ref{fig:ap_usage} depicts active subscribers and hourly throughput of the network.

Network policies for the \aparks networks are very simple: because the LTE network simply serves as backhaul, all UEs simply have unrestricted access.
Per-user policies are implemented by \aparks's pre-existing captive portal and pre-paid billing software, which is implemented using standard techniques (i.e., RADIUS for AAA at the WiFi AP).

\textbf{Operational complexity.} 
\aparks's original \name pilot was motivated in part by their poor experiences with the operational complexity of other commercial and open-source cellular core software in their previous two years of deployment.
Although operational complexity is subjective, one quantifiable way in which it manifests is in an operator's labor costs: simpler systems should require less staff time and support to manage.
Table~\ref{t:aparks-cost} shows the results of this comparison for \aparks.
For identical access network infrastructure, \aparks achieved a 43\% reduction in per-site deployment costs using \name compared to traditional architectures, largely driven by a reduction in support costs and engineering time for site configuration and planning.\footnote{Unfortunately, we do not have data on ongoing maintenance costs from \aparks; however, \aparks' decision to use \name for future deployments suggests it compared favorably.}

\begingroup
\setlength{\tabcolsep}{2pt} %
\renewcommand{\arraystretch}{1} %
\begin{table}[t]
\footnotesize
    \begin{center}
        \ra{1.06}
        \begin{tabular}{@{}lrrcp{2.5cm}@{}} \toprule
            \textbf{Item} & \textbf{Traditional} & \textbf{\name} & \textbf{Difference (\%)} & \textbf{Notes} \\ \hline
            RAN & \$7,950 & \$7,950 & - & Identical RAN and backup power.\\
            Core HW & \$1,200 & \$300 & -\$900 (-75\%) & \\
            Core SW & \$2,000 & \$600 & -\$1,400 (-70\%) & Licenses/support. \\
            Field Eng. & \$200 & \$200 & - & Installation. \\
            LTE Eng. & \$5,000 & \$330 & -\$4,670 (-93\%) & Planning, core config. \\
            \hline
            \textbf{Cost/Site} & \$16,350 & \$9,380 & -\$6,970 (-43\%) &  \\ 
            \bottomrule
        \end{tabular}
    \end{center}
    \vspace{-15pt}
    \caption{Comparison of per-site installed costs for \aparks's traditional cellular system compared to Magma. Total cost per site decreased by 43\%, driven primarily by \name's reduction in operational complexity for deployment.}
    \label{t:aparks-cost}
    \vspace{-0.2in}
\end{table}
\endgroup

\subsubsection{Franchised MNO Extension}\label{sec:eval_fed}
A second (and, to our knowledge, the largest) deployment of \name is an early-stage deployment to provide a franchised, neutral host network.\footnote{A neutral host network describes a business model in which a mobile network is operated by an entity for the sole purpose of providing wholesale capacity to third-party retail MNOs and MVNOs; the neutral host network operator does not have its own users, but instead enables users of its customers to use the neutral host network on a shared basis.}
This network is unique in that the physical deployment of network infrastructure is not managed by any single network operator.
Instead, ``micro network operators'' (which include individuals, small ISPs, and enterprises) deploy LTE and 5G RAN equipment alongside \name AGWs that have been customized by \ffi to support their proprietary traffic accounting and settlement system.

\textbf{Services and Policy.}
The neutral host network is operated by \ffi and allows customers of incumbent MNOs to use this network for service.
The core ``policy'' supported by this network is tunnelling all user traffic back to the appropriate MNO; a user's MNO, in turn, applies their standard network policies for billing, charging, and throttling within their existing core network.
The \ffi network provides access on a best-effort basis, with each micro network operator leveraging shared CBRS~\cite{cbrs} spectrum in the 3.5GHz band (as done in the previous deployment).
This service requires integrating the thousands of distributed AGWs with a partner MNO's centralized core network, leveraging the federation capabilities described in Section~\ref{sec:federate}.

\textbf{Scale.}
As of this writing, this network is still in early testing, so does not have significant user traffic.
However, it still provides a useful example of how the \name \emph{control plane} scales with network size: even without users, \name still manages device configuration, network monitoring, and supports interconnection with partner MNO core networks.

The \ffi network began initial deployments in November 2021, and as of April 2022 consists of 5370 AGWs and 880 eNodeBs (\ffi reports the discrepancy between AGWs and eNodeBs is due to supply-chain issues: while AGWs are commodity x86 PCs, cellular radios are specialized equipment with fewer vendors and the ones used in this network only began shipping in January 2022).
The network is currently adding on average 150 new AGWs and 90 new eNodeBs per week, all of which are deployed on an ad-hoc basis by micro-network operators; these AGWs are deployed in 45 states across the United States.\footnote{The network only operates in the United States for regulatory reasons.}

Supporting this network is a dedicated \orctr running on six AWS virtual machines managed by Kubernetes (EKS)~\cite{eks}.
Three instances are dedicated towards ``heavy'' tasks: operation of the FeG, device configuration, and metrics reporting; these systems are each equipped with 16 vCPUs and 32GB RAM.
Remaining \orctr services run on a collection of smaller VMs (4 vCPU/16GB RAM).
The GTP-A runs on a single bare metal server with a 3.4GHz 8-core Xeon E2278G CPU, 32GB RAM, and 2x10G NICs, and is physically co-located near the facilities of a partner MNO's core network. 
In total, this costs \ffi approximately \$4,000 per month to operate.

We view the rapid deployment of this network as cautious evidence for \name's ability to support large-scale networks with unique business models.
We hope to further investigate the operational dynamics of this network in future work.

\section{Related Work}
\label{sec:related}

\textbf{Open-source LTE/5G core networks:} 
Several projects share our goal of creating an open-source LTE/5G cellular core network~\cite{www-free5gc,open5gs,www-oai,www-srsran};
these were preceded by similar efforts to build open 2G and 3G networks~\cite{openbts,osmocom}.
With the exception of OpenBTS~\cite{openbts} (a GSM-to-VOIP bridge), each of these focuses on implementing traditional, 3GPP-compliant, core networks.\footnote{We note that the \name AGW's LTE-specific portion was originally based upon OpenAirInterface~\cite{www-oai}, as it was the most mature open-source core available at the inception of \name's development.}
Aether~\cite{www-aether,video-aether} is an open-source 5G-connected edge platform,
which brings together 5G connectivity and edge-cloud servers.  Like Magma, Aether adopts cloud design principles. However, Aether does not refactor the network design to break the coupling of the radio access technology with the core, and Aether does not focus on low-cost equipment to reach under-served users.

\textbf{Expanding connectivity access:}
Many efforts have proposed or described novel solutions for expanding Internet access to under-served people~\cite{wildnet,kiosknet,daknet,dye2018paquete,waites2016remix,potsch2018zyxt,surana2008beyond}.
Similarly, small(er)-scale network operators have a rich history providing service to especially rural communities~\cite{microtelco}, such as community networks~\cite{guifi, nycmesh,altermundi,belli2017community} and small ISPs~\cite{hasan2015challenges}.
Of this extensive literature, \name is most closely related to work on community cellular networks~\cite{cn_ictd13, anand2012villagecell,rhizomatica}.

\textbf{NextG cellular core architecture.}
The networking research community is actively rethinking the design of next generation networks.
PEPC~\cite{qazi2017pepc} refactors the packet core by consolidating user state into one location, similar in spirit to Magma's AGW.
ECHO~\cite{nguyen2018echo} refactors an EPC to run on less-reliable public cloud infrastructure.
SCALE~\cite{banerjee2015scale} explores an elastically scalable cellular control plane, and KLEIN~\cite{qazi2016klein} describes a similarly elastic control and data plane.
Although these works all focus on (logically) centralized core networks, the techniques described are complementary to \name.

Other work takes a more ``clean slate'' approach to reimagining the cellular core.
CellBricks~\cite{cellbricks} contemplates a highly federated cellular network and moves support for mobility, authentication, and billing into end hosts; it is implemented as an extension to \name.
dLTE~\cite{johnson2018dlte} makes 4G networks more like WiFi through a decentralized design, including a global registry for peer discovery.
SoftCell~\cite{jin2013softcell} uses SDN principles to improve the scalability and flexibility of the packet core network.
\name draws on this body of work for inspiration while maintaining a backwards-compatible, standards-compliant edge to facilitate production deployment.

\name directly builds on recent work exploring core architectures for under-served communities.
CCM~\cite{hasan2019ccm} presents a distributed cellular 2G core that enables semi-disconnected operation over unreliable rural backhaul connections; this work served as an early inspiration for \name, which extends these concepts to modern wireless access technologies.
Similarly, CoLTE~\cite{sevilla2019colte} provides a lightweight core which---like an AGW---is co-located with RAN elements, but unlike \name focuses on small, independent community networks.

\textbf{Open radio access networks:} 
Several recent initiatives focus on opening up the radio access network (RAN).  For example,
the OpenRAN project~\cite{openran} and the O-RAN alliance~\cite{oran,www-oran} develops standards that disaggregate 3GPP RANs, with open interfaces between the layers.
These efforts are complementary to Magma, as they focus on the cellular interface---the part of the network \emph{before} reaching Magma's access gateway.

\section{Conclusion} 
\label{sec:concl}
We have presented our experiences in designing
and deploying \name, an open-source platform for building access networks.
The most important design decision was to terminate the RAN-specific protocols 
in access gateways close to the radio.
This simple design decision brings many benefits: supporting diverse radio technologies, tolerating disruptions in backhaul links, using a low-cost software data plane, and scaling naturally with a hierarchical SDN control plane. Magma also adopts modern cloud-computing design patterns (e.g., desired-state synchronization, tolerance to failure of individual components) and open-source software components (e.g., gRPC, Open vSwitch, Kubernetes, Prometheus).
In line with \name's goal to enable practical networks, we demonstrated that \name can support typical deployment scenarios and discussed two large-scale commercial networks that use \name. Importantly, \name also scales {\em down}, with a small minimum footprint that supports incremental deployment, thus filling a gap between traditional WiFi and cellular.
All software artifacts for \name are available on GitHub\footnote{\url{https://github.com/magma/magma}}.

Magma was designed with the primary goal of reaching under-served communities, by supporting heterogeneous radio and backhaul technologies and reducing capital and operational cost.  We believe that Magma is a good fit for other deployment scenarios, including enterprise 5G networks. Future work on Magma can expand the set of supported features, including seamless mobility between access gateways as well as network virtualization. We look forward to extending the Magma code base, and the community of contributors to the software, so the platform can evolve to serve more users.

\section*{Acknowledgements} 
\label{sec:ack}
We thank our shepherd Ranveer Chandra and the anonymous reviewers for their helpful feedback. We thank Boris Renski, Matthew Mosesohn, and Joey Padden for their assistance gathering deployment data for this paper. We also thank the Magma developer and user community for their important contributions, as well as Meta Connectivity for supporting the early development and deployments of Magma.

\bibliographystyle{plain}
\bibliography{magma}
\newpage\section*{Appendix}

\begingroup
\setlength{\tabcolsep}{4pt} %
\renewcommand{\arraystretch}{1} %
\begin{table}[bh]
\scriptsize
    \centering
    \begin{tabular}{l|l} \toprule
        \textbf{Acronym}  & \textbf{Definition}\\ \hline
        MME & Mobility Management Entity\\
        HSS & Home Subscriber Server\\
        PCRF & Policy and Charging Rules Function\\
        SGW & Serving Gateway\\
        PGW & Packet Gateway\\
        AMF & Access and Mobility Function\\
        SMF & Session Management Function\\
        PCF & Policy Control Function\\
        UDM & Unified Data Management\\
        AUSF & Authentication Server Function\\
        S1AP & S1 Access Protocol\\
        NGAP & Next Generation Access Protocol\\
        SCTP & Stream Control Transmission Protocol\\
        NAS & Non-Access Stratum\\
        RAN & Radio Access Network\\
        LTE & Long Term Evolution\\
        3GPP & Third Generation Partnership Project\\
        UE & User Equipment (a phone or other cellular client)\\
        eNodeB & The ``access point'' for an LTE network\\
        gNodeB & The ``access point'' for an 5G network\\
        AGW & Access Gateway\\
        AAA & Authentication, Authorization, and Accounting\\
        RADIUS & Remote Authentication Dial-In User Service\\
    \end{tabular}
    \caption{Acronyms used in the paper}
    \label{tab:acronym}
    \vspace{-0.1in}
\end{table}
\endgroup

\end{document}